\documentclass[twocolumn]{aastex62}

\newcommand {\kms}{$\rm{km\ s^{-1}}$}
\newcommand{\ergcms}{$\rm ~erg~cm^{-2}~s^{-1}$}

\received{December 21, 2018}
\revised{\today}
\accepted{\today}
\submitjournal{ApJ}

\shorttitle{A 3D view of H$_2$ in SN 1987A}
\shortauthors{Larsson et al.}

\begin{document}

\title{A 3D view of molecular hydrogen in Supernova 1987A}

\correspondingauthor{J. Larsson}
\email{josla@kth.se}

\author[0000-0003-0065-2933]{J.Larsson}
\affil{Department of Physics, KTH Royal Institute of Technology, The Oskar Klein Centre, AlbaNova, SE-106 91 Stockholm, Sweden}

\author[0000-0001-6815-4055]{J. Spyromilio}
\affiliation{European Southern Observatory, Karl-Schwarzschild-Strasse 2, D-85748 Garching, Germany}

\author[0000-0001-8532-3594]{C. Fransson}
\affiliation{Department of Astronomy, Stockholm University, The Oskar Klein Centre, AlbaNova, SE-106 91 Stockholm, Sweden}


\author[0000-0002-4663-6827]{R. Indebetouw}
\affiliation{National Radio Astronomy Observatory and University of Virginia, 520 Edgemont Road, Charlottesville, VA 22903, USA}

\author[0000-0002-5529-5593]{M. Matsuura}
\affiliation{School of Physics and Astrophysics, Cardiff University, Queens buildings, The Parade, Cardiff CF24 3AA, UK}

\author[0000-0002-5724-1636]{F. J. Abell\'{a}n}
\affiliation{Departamento de Astronomía y Astrofísica, Universidad de Valencia, C/Dr. Moliner 50, E-46100 Burjassot, Spain}

\author[0000-0002-8736-2463]{P. Cigan}
\affiliation{School of Physics and Astrophysics, Cardiff University, Queens buildings, The Parade, Cardiff CF24 3AA, UK}

\author[0000-0003-3398-0052]{H. Gomez}
\affiliation{School of Physics and Astrophysics, Cardiff University, Queens buildings, The Parade, Cardiff CF24 3AA, UK}

\author[0000-0002-4413-7733]{B. Leibundgut}
\affiliation{European Southern Observatory, Karl-Schwarzschild-Strasse 2, D-85748 Garching, Germany}

\begin{abstract}

Supernova (SN)~1987A is the only young SN in which H$_2$ has been detected in the ejecta. The properties of the H$_2$ are important for understanding the explosion and the ejecta chemistry. Here, we present new VLT/SINFONI observations of H$_2$ in SN 1987A, focussing on the $2.12\ \mu$m (1,0)S(1) line. We find that the 3D emissivity is dominated by a single clump in the southern ejecta, with weaker emission being present in the north along the plane of the circumstellar ring. The lowest observed velocities are in the range 400--800~\kms, in agreement with previous limits on inward mixing of H. The brightest regions of H$_2$ coincide with faint regions of H$\alpha$, which can be explained by H$\alpha$ being powered by X-ray emission from the ring, while the H$_2$ is powered by  $^{44}$Ti. A comparison with ALMA observations of other molecules and dust shows that the brightest regions of H$_2$, CO and SiO occupy different parts of the inner ejecta and that the brightest H$_2$ clump coincides with a region of very weak dust emission. The latter is consistent with theoretical predictions that the  H$_2$ should form in the gas phase rather than on dust grains. 


\end{abstract}


\section{Introduction}
\label{sec:intro}

The proximity of SN~1987A has allowed for a large number of unique observations. One of these is the detection of H$_2$ in the ejecta \citep{Fransson2016}, which confirmed theoretical predictions \citep{Culhane1995}. Specifically, observations with VLT/SINFONI revealed rovibrational lines in the near-IR (NIR)  at  $2.12\ \mu$m and $2.41\ \mu$m. The former is identified with the (1,0)S(1) transition, while the latter is a blend of the (1,0)Q(1--3) transitions at 2.406, 2.413, and 2.423~$\mu$m. 

The H$_2$ lines were detected in all four SINFONI K-band observations carried out between 2005--2014. No significant time-evolution in flux was detected and the morphology was observed to be centrally concentrated. This shows that the energy source is likely to be $^{44}$Ti rather than X-ray emission from the circumstellar ring that surrounds the ejecta.  The molecules may be excited either by UV fluorescence or non-thermal electrons. While the modeling in \cite{Fransson2016} slightly favored the UV-scenario, no clear conclusion could be drawn.

Observing H$_2$ adds to previous observations of other molecules in SN~1987A. Both CO and SiO were observed in the IR within months after the explosion \citep{Spyromilio1988,Atiken1988,Rank1988}. Much more recently, ALMA has detected rotational transitions in the millimeter range from CO, SiO, HCO$^{+}$ and SO \citep{Kamenetzky2013,Matsuura2017}. The two strongest transitions, CO~$\rm{J} = 2-1$ and SiO~$\rm{J} = 5-4$, have also been studied in 3D with 50 mas angular resolution \citep{Abellan2017}. It is possible to obtain 3D information about the explosion geometry in SN~1987A because the inner ejecta are spatially resolved and expanding homologously. The homologous expansion, which was reached about a week after the explosion \citep{Gawryszczak2010}, allow us to use Doppler shifts to determine the distance along the line of sight from the centre of the explosion. This technique has previously been used to study the 3D emissivities of a number of atomic lines in the ejecta (\citealt{Kjaer2010, Larsson2013,Larsson2016}, L16 from hereon). 

The asymmetric 3D morphology of the freely expanding ejecta is interesting because it carries information about the explosion mechanism and the nature of the progenitor (e.g., \citealt{Wongwathanarat2015}). In the case of SN~1987A, the circumstellar ring acts as a useful reference point for studying the morphology. The ring has a diameter of $\sim 1.2$ light years ($1\farcs{6}$) and is inclined by $44^{\circ}$  to the line of sight \citep{Plait1995}. It was formed about 20,000 years before the explosion \citep{Crotts1991} and while it is not clear how it formed, it most likely defines the equator of the progenitor star. The 3D maps of CO and SiO reveal torus/shell-like distributions perpendicular to the ring for the brightest emission, accompanied by faint extended tails towards the south \citep{Abellan2017}. This is different from the atomic lines (where the best 3D data is available for H$\alpha$ and the [Si~{\scriptsize I}]+[Fe~{\scriptsize II}] line at 1.644~$\mu$m), which show a clear north/south asymmetry. Specifically, the emission is distributed between the line of sight and the plane of the ring in the north, but closer to the plane of the sky in the south

Here, we present new SINFONI K-band observations of the H$_2$ emission in SN 1987A obtained in December 2017 and January 2018. The use of a laser guide star (which improves the spatial resolution), together with a long exposure time and the continuously expanding ejecta,  allow us to study the 3D morphology of the H$_2$. The expansion means that the ejecta are now 65\% larger on the sky as compared to our first observations in 2005. We also compare the 3D distribution with the previous 3D maps of H$\alpha$, CO and SiO. Finally, we use our previous observations to study the time-evolution of H$_2$. 

This paper is organized as follows: after describing the observations in Section \ref{sec:obs}, Section \ref{sec:h2results} presents the spectra, images and 3D morphology of H$_2$. Section \ref{sec:comp} contains a comparison with the 3D morphology of H$\alpha$ and the ALMA observations of molecules and dust, while Section \ref{sec:time} deals with the time-evolution of H$_2$. The discussion and conclusions are presented in Sections \ref{sec:disc} and \ref{sec:conclusions}, respectively. A study of the continuum emission in the K-band is presented in Appendix \ref{app:cont}. 

Throughout this paper we assume a distance to SN~1987A in the LMC of 51.2~kpc \citep{Panagia1991} and that the centre of  the ring (determined by \citealt{Alp2018}) coincides with the centre of the explosion. All uncertainties are 1$\sigma$  unless otherwise stated.

\section{Observations and data reduction}
\label{sec:obs}

\begin{deluxetable*}{cccccc}[t]
\tablecaption{Observations \label{tab:compobs}}
\tablecolumns{5}
\tablenum{1}
\tablewidth{0pt}
\tablehead{
\colhead{Instrument} &
\colhead{Filter/Grating} &
\colhead{Line/Emission component} &
\colhead{Date(s)} & 
\colhead{Epoch\tablenotemark{a}} & 
\colhead{Reference} \\
\colhead{} & 
\colhead{} &
\colhead{} &
\colhead{(YYYY-mm-dd)} &
\colhead{(d)} &
\colhead{}
}
\startdata
VLT/SINFONI &  K  & H$_2$  & 2017-12-11 -- 2018-01-30 & 11,275 & \nodata \\
VLT/SINFONI &  K  & H$_2$  & 2014-10-12 -- 2014-12-01 & 10,120 & 1,2 \\
VLT/SINFONI &  K  & H$_2$  & 2010-11-05 -- 2011-01-02 & 8,694 & 2,3 \\
VLT/SINFONI &  K  & H$_2$  & 2007-11-07 -- 2080-01-19 & 7,615 & 2 \\
VLT/SINFONI & K  & H$_2$ & 2005-10-30 -- 2005-11-14 & 6,832 & 2,3,4 \\
HST/WFC3 & F625W & mainly H$\alpha$ & 2018-07-08 & 11,458 & \nodata \\
HST/STIS & G750L & H$\alpha$ & 2014-08-16 -- 2014-08-20 & 10,037 & 1  \\
ALMA & Band 7 &  dust (311.9--319.5~GHz) & 2015-06-28 -- 2015-09-22 & 10,395 & 5 \\
ALMA & Band 6 & CO 2--1, SiO 5--4& 2014-09-02, 2015-11-02 & 10,053, 10,479 & 6 \\
\enddata
\tablenotetext{a}{Days since explosion on 1987-02-23. An exposure-weighted mean is used when the observations were spread over an extended period.}
\tablecomments{All observations provide 3D information apart from the HST/WFC3 and ALMA/315~GHz images.}
\tablerefs{(1) L16; (2) \cite{Fransson2016}; (3) \cite{Larsson2013}; (4)  \cite{Kjaer2010}; (5) P. Cigan et al. (in preparation), (6)  \cite{Abellan2017}}
\vspace{-0.5cm}
\end{deluxetable*}

\subsection{SINFONI observations in 2017/2018}

Observations by the SINFONI Integral Field Unit \citep{Bonnet2003,Eisenhauer2003} at the VLT were carried out between 2017-12-11 and 2018-01-30 as a part of ESO Program 0100.D-0705(C) (Table~\ref{tab:compobs}). The dates correspond to an exposure-weighted epoch of 11,275 days after explosion. We will simply refer to the date of these observations as 2017 from here on. The total exposure time in the K-band was 10,800~s.  The data were acquired in 600~s exposures taken in ABBA cycles.  The standard ESO pipeline \citep{Modigliani2007} was used to generate intermediate data cubes for each of the nodded frames. The ESO pipeline corrects for distortion, wavelength calibrates and also performs a first sky subtraction in each pair. We then combine the intermediate data cubes using our own software for sub-pixel alignment, combination and sky-subtraction. This removes outliers at the sub-pixel scale before rebinning the data to the original resolution.

For the flux calibration we use the standard spectrum from the concurrent observations. Based  on the catalogued spectral energy distribution we then generate a response spectrum that is multiplied into the final data cubes. With multiple standard stars observed on any given SINFONI run it is possible to check the consistency of the fluxing solution. We find that the instrumental response is remarkably stable, with the response curves differing by $<10\%$. 

The point-spread function (PSF) in SINFONI has an enhanced core and broad wings, which means that it is not well modeled by a single Gaussian. To assess the spatial resolution we instead determine the  $50\%$ ($80\%$) encircled energy, which is $0\farcs{10}$ ($0\farcs{18}$).  The broad wings of the PSF means that spectra extracted from the ejecta region are contaminated by emission from the bright ring. To correct for this scattered light we use the fact that the lines from the ring are much narrower than the lines from the ejecta (full width half maximum, FWHM,  $\sim 300$ and $3000$~\kms, respectively). The narrow components can be clearly identified due to the good spectral resolution of 70~\kms. This means that the scattered component can be removed by subtracting the ring spectrum multiplied with a constant from the ejecta spectrum. The constant is chosen such that the narrow components disappear. An example of this correction for SINFONI data is provided in L16 (their Fig.~7). The uncertainty in this correction does not significantly affect our results since there are no strong lines from the ring overlapping with the H$_2$ lines from the ejecta (cf. Fig.~\ref{fig:kspec}). All spectra were also corrected for the systematic velocity of SN~1987A of $287$~\kms \citep{Groningsson2008a}.

\subsection{Other observations used for comparison}

Observations used for comparison with other emission components in the ejecta and for studying the time-evolution of H$_2$ are listed in Table~\ref{tab:compobs}. All observations apart from the HST/WFC3 and ALMA 315~GHz (dust) images provide 3D information and have been analyzed in previous works. The previous VLT/SINFONI K-band observations have been used to study the H$_2$ emission \citep{Fransson2016} and the bright He~{\scriptsize I} 2.058~$\mu$m line (\citealt{Kjaer2010,Larsson2013}; L16). The HST/STIS data were used to study a number of optical emission lines (H$\alpha$, [Ca~{\scriptsize II}]$\ \lambda \lambda 7292,\ 7324$, [O~{\scriptsize I}]$\ \lambda \lambda 6300,\ 6364$ and Mg~{\scriptsize II}$\ \lambda \lambda 9218,\ 9244$; L16), while the ALMA/Band~6 observations were used to study molecular lines (CO~2--1 and SiO~5--4; \citealt{Abellan2017}).  

The HST/WFC3 image was obtained in the F625W filter using four $300\ \rm{s}$ dithered exposures. These were combined using DrizzlePac\footnote{\url{http://drizzlepac.stsci.edu}}, which removes cosmic rays, applies distortion corrections and improves the spatial resolution using the drizzle technique \citep{Gonzaga2012}. 

The dust image from ALMA is based on four 19~min observations in Band 7 acquired as a part of the  ALMA project 2013.1.00063 (PI Indebetouw).  The data were reduced with the Common Astronomy Software Application package (CASA\footnote{\url{http://casa.nrao.edu/}}, \citealt{McMullin2007}) version 4.5.1. The continuum image was constructed from a narrow frequency range around 315~GHz, where no lines are expected according to the spectral model in \cite{Matsuura2017}. We note that this is different from using the ALMA continuum setting. The beam size was 0\farcs19$\times$0\farcs14 with a position angle of $29^{\circ}$ in anti-clockwise direction from the north. The details of the observations and data reduction are described by P. Cigan et al. (in preparation). 

The different data sets were aligned using the positions of nearby stars or the ring of SN~1987A, with the exception of the ALMA data. For ALMA we instead use the absolute astrometry together with the position of SN~1987A \citep{Alp2018}. The accuracy of the ALMA absolute astrometry is $\lesssim 0\farcs{01}$, while the accuracy of the alignment of the other data sets is $< 0\farcs{02}$ (see \citealt{Alp2018} for details).

\section{H$_2$ emission at the end of 2017}
\label{sec:h2results}

\subsection{spectra and images}
\label{sec:specim}
\begin{figure*}[t]
\plotone{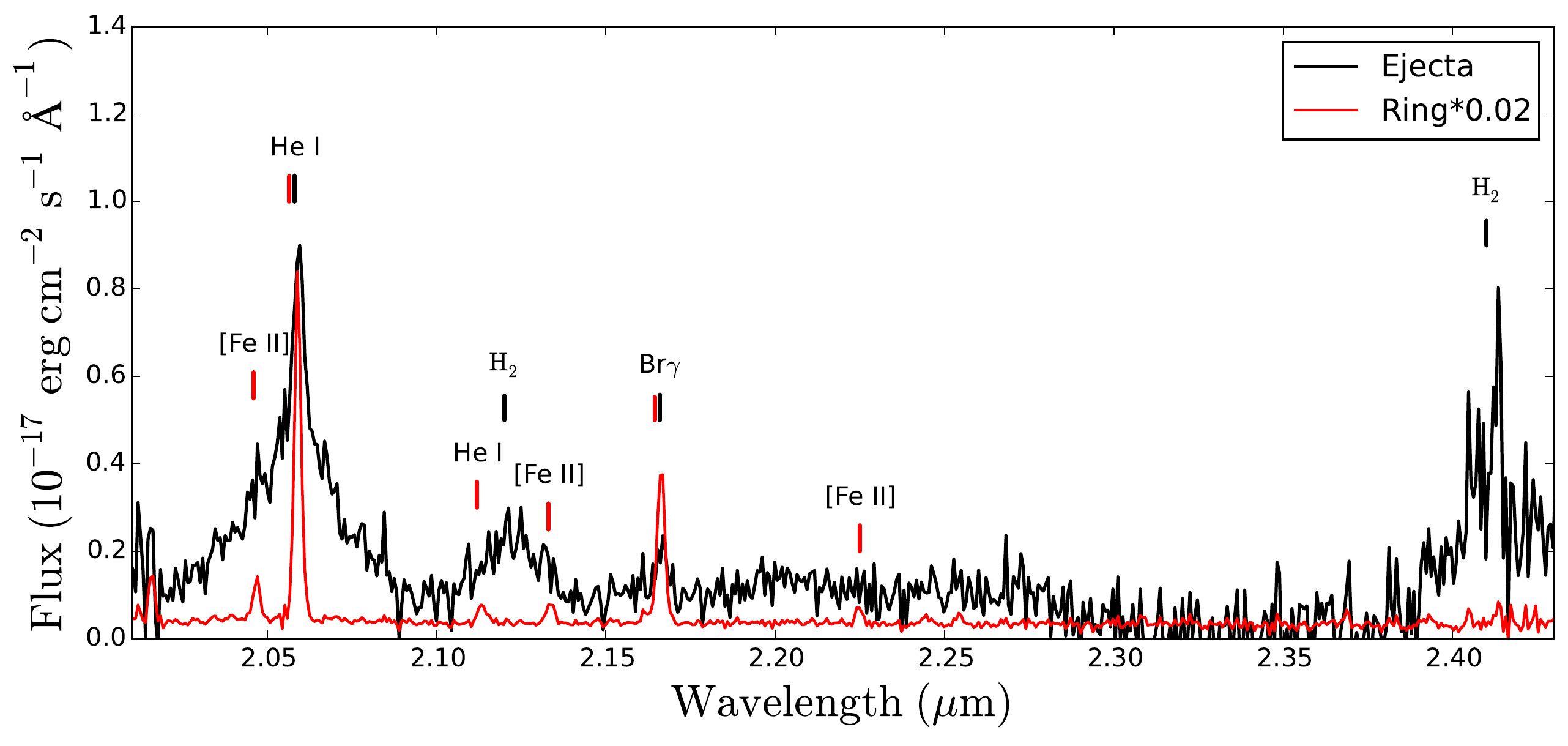}
\caption{SINFONI K-band spectra from 2017 of the ejecta (black) and ring (red). The ring spectrum has been rescaled by a factor of 0.02 in order to allow comparison with the ejecta spectrum. Both spectra have been binned by a factor of three for visual clarity. The identification of atomic lines is from \cite{Kjaer2007}. We only label lines that are strong enough to be clearly detected, with black and red bars indicating lines in the ejecta and ring spectra, respectively. \\ \label{fig:kspec}}
\end{figure*}
\begin{figure*}[t]
\plottwo{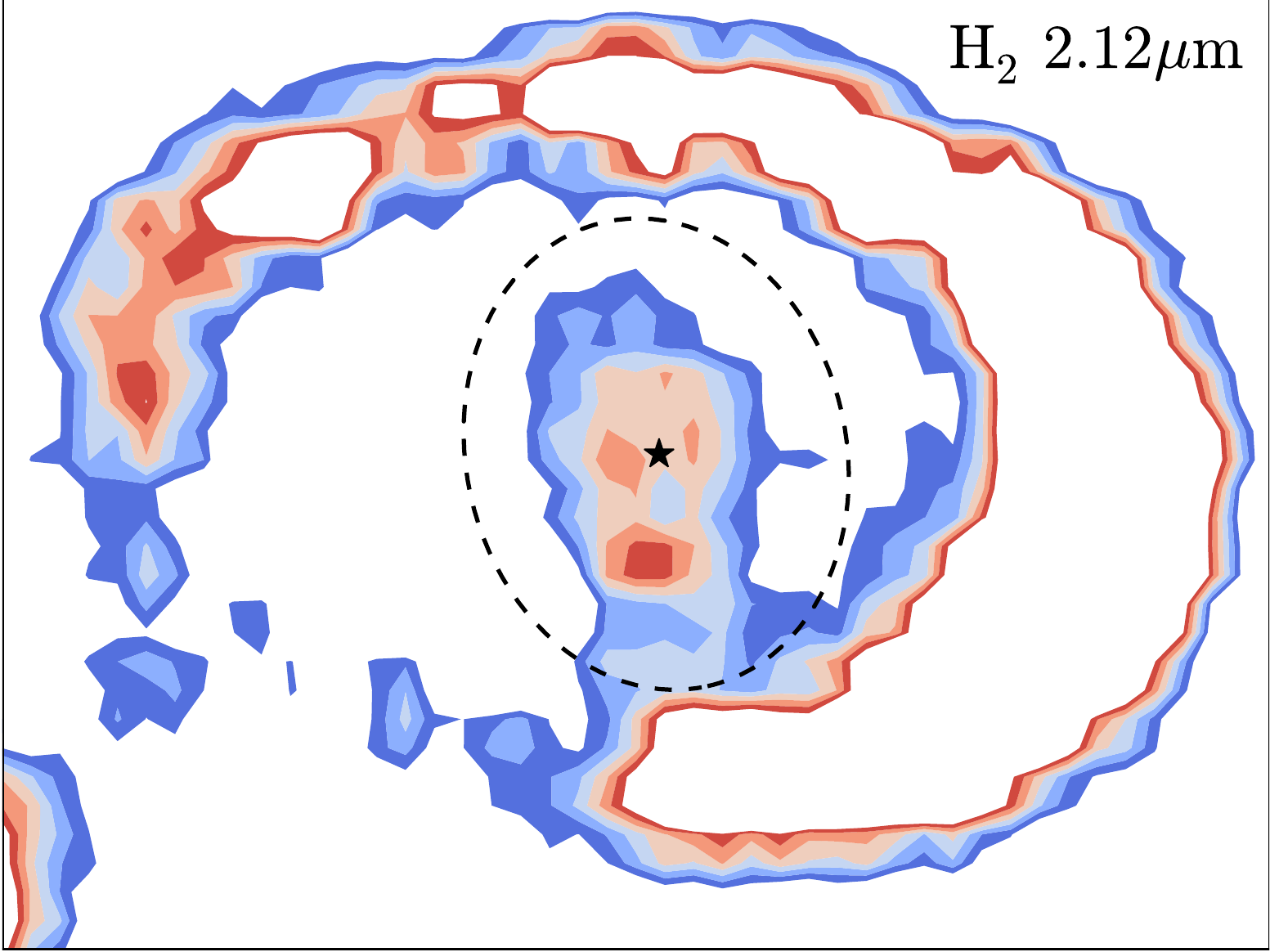} {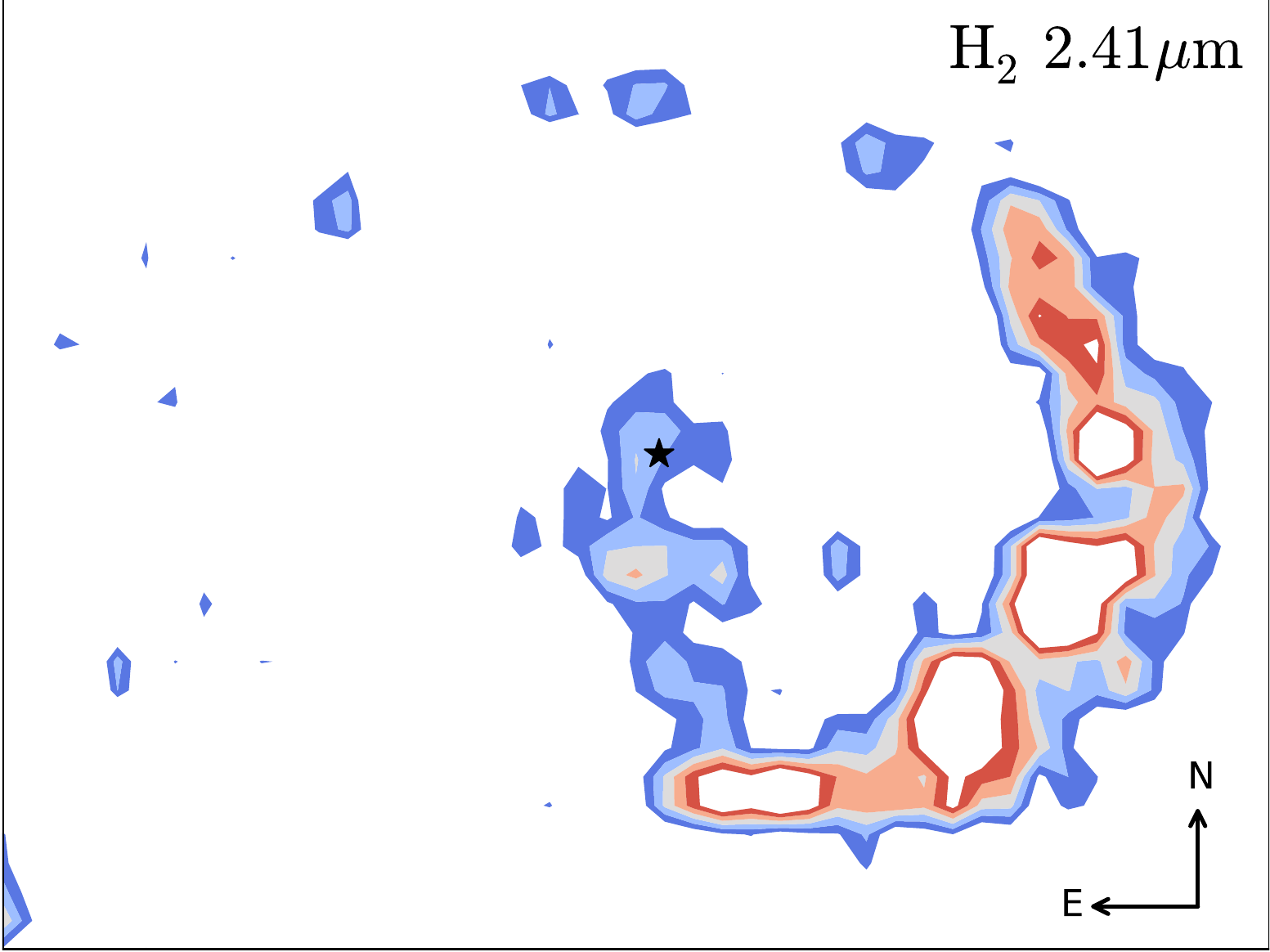}
\caption{Contour plots of the H$_2$ lines at $2.12\ \mu$m (left) and $2.41\ \mu$m (right) produced from the SINFONI K-band observation in 2017. The interval $\pm 2000$~\kms\ around each line is included. The first contour is 3$\sigma$ above the background level inside the ring and subsequent contours are separated by one $\sigma$.  The black star marks the center of the ring and the black ellipse in the left panel shows the aperture used for extracting spectra. The FOV is $2\farcs{2} \times 1\farcs{7}$. \\ \label{fig:h2conts}}
\end{figure*}

The K-band spectrum of the ejecta in 2017 is shown in Fig.~\ref{fig:kspec}. The spectrum was extracted from an elliptical aperture with semi-major axis  $0\farcs{41}$, a ratio of minor to major axis of 0.8 and a position angle of $12^{\circ}$ (defined anti-clockwise from north), which covers the full ejecta while avoiding the ring (see Fig.~\ref{fig:h2conts}). The spectrum of the ring, which has much narrower lines, is also shown in Fig.~\ref{fig:kspec} for comparison. The strongest lines from the ejecta are the He~I line at $2.058\ \mu$m, followed by the two H$_2$ lines at $2.12$ and $2.41\ \mu$m, respectively. There are no H$_2$ lines present in the ring spectrum. The H$_2$ line at $2.41\ \mu$m has very low signal-to-noise (S/N) since it is at the edge of the filter where effects of the atmosphere are strong. However, this line still provides a clear identification of H$_2$ from the ejecta since it is not blended with any other lines or continuum emission.  

The H$_2$ $2.12\ \mu$m line has much better S/N, but is located in a region of the spectrum where there is an additional smooth emission component.  We will refer to this component as the ``continuum", although we note that it is likely composed of weak lines and the tails of nearby broad lines, as discussed in Appendix \ref{app:cont}.  According to the model presented in \cite{Kjaer2010}, the strongest lines directly overlapping with H$_2$ are the  Al~{\scriptsize I} 2.1093 2.1163~$\mu$m transitions. These lines are expected to be at least a factor 2 weaker than the blue wing of the H$_2$ $2.12$~$\mu$m line at 2.11~$\mu$m.  

Fig~\ref{fig:h2conts} shows contour plots of the two H$_2$ lines inside $\pm 2000$~\kms. The contours are separated by one standard deviation, as measured from source-free regions in the images ($1.7\times 10^{-19}$ and $5.3\times 10^{-19}$~\ergcms\ for 2.12 and 2.41~$\mu$m, respectively). The first contour is 3$\sigma$ above the background level inside the ring. The emission from the ring in these images is due to H~{\scriptsize I} continuum, as well as emission lines from He~{\scriptsize I} and Fe~{\scriptsize II} in the case of the 2.12~$\mu$m image. The morphology of the ejecta is dominated by a bright clump in the south, clearly seen in both H$_2$ lines.  The 2.12~$\mu$m image, which has much better quality, also shows significant emission in a more extended region around the center. 

The results in the following sections will focus on the 2.12~$\mu$m line due to its better S/N. However, we stress the importance of the 2.41~$\mu$m line in confirming that the bright clump in the south is due to H$_2$ emission, showing that this feature is not significantly affected by the continuum. Since the lines making up the continuum are most likely powered by X-ray emission from the ring (see Appendix~\ref{app:cont}), we expect the continuum to contribute to low-level emission in the south-west in the images of the $2.12\ \mu$m line in 2017. This may explain part of the faintest emission in the southern ejecta in the  $2.12\ \mu$m line in Fig.~\ref{fig:h2conts}, but not all of it since the 2.41~$\mu$m line also shows emission in the south. The 3D plots presented in the following section are not affected by the continuum since we only consider iso-surfaces well above the continuum level.

\subsection{3D morphology}
\label{sec:h23d}

\begin{figure}[t]
\resizebox{90mm}{!}{\includegraphics{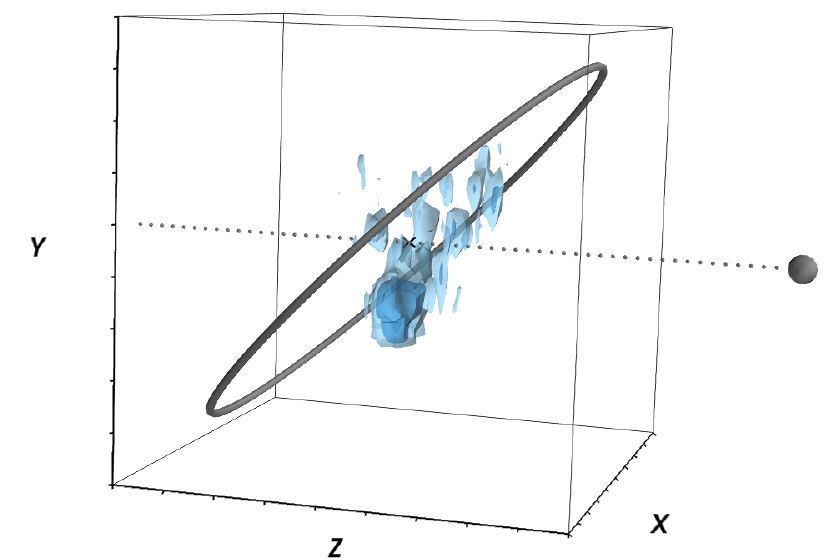}}
\caption{3D iso-surfaces of H$_2$ corresponding to 45$\%$ (light blue) and 65$\%$ (dark blue) of the maximal intensity. The ring indicates the position of the reverse shock at the inner edge of the ring.  The centre of the ring is marked by a cross. The dashed line and blob show the line-of-sight and position of the observer, respectively. The tick marks on the coordinate axis are separated by $1000$~\kms. An animated (rotating) version of this figure is available online.  \label{fig:h2_3d}}
\end{figure}

Since the inner ejecta of SN~1987A are expanding freely, the distance from the centre of the explosion is directly proportional to the velocity of the ejecta. We can therefore use the SINFONI IFU data to obtain a 3D view of the emissivity of H$_2$. At the time of the SINFONI observation in 2017, $0\farcs{1}$ corresponds to 790~\kms. When presenting the results it is natural to use \kms\  as a unit of distance within the ejecta. This also allows for direct comparisons with 3D observations from different epochs (Section \ref{sec:comp}) even though the linear dimensions are changing due to the expansion. 

Fig.~\ref{fig:h2_3d} shows 3D iso-surfaces corresponding to 45$\%$ and $65\%$ of the maximal value after subtraction of the continuum (calculated in the range $2500-3500$~\kms). The cube has been binned by a factor of six in the spectral direction to improve the S/N, resulting in bins of 210~\kms\  along the line of sight. An intensity level of $\sim 30\%$ of max corresponds to 3$\sigma$ above the background level, which means that the plotted contour levels represent highly significant emission. The 3D map shows that the emission is dominated by the bright clump in the south (also seen in the 2D contour plots in Fig.~\ref{fig:h2conts}). The centre of the clump has a space velocity of  $\sim 1700$~\kms, directed towards the south and away from the observer. Along the line of sight (where the resolution is best) the centre is redshifted by $\sim 400$~\kms\  and the extent is $\rm{FWHM} \sim 1400$~\kms. To the north of this clump, there is weaker emission 
in a fragmented structure roughly in the plane of the ring. 

The fraction of the total volume inside 2000~\kms\ occupied by emission brighter than the the plotted $45\%$ and $65\%$ contour levels are 0.12 and 0.03, respectively. The 3D geometry shows a gradual transition to increasingly more redshifted emission when going from north to south, as previously seen in several other emission lines from the ejecta (L16). The ratio of blue- to redshifted emission in the three $0\farcs{5} \times 0\farcs{25}$ regions defined in L16 are $1.02\pm 0.14$ (north), $0.68\pm 0.06$ (middle) and $0.07\pm 0.05$ (south).

\section{Comparison with other ejecta emission components}
\label{sec:comp}

\subsection{Comparison with H$\alpha$}

\begin{figure}[t]
\resizebox{90mm}{!}{\includegraphics{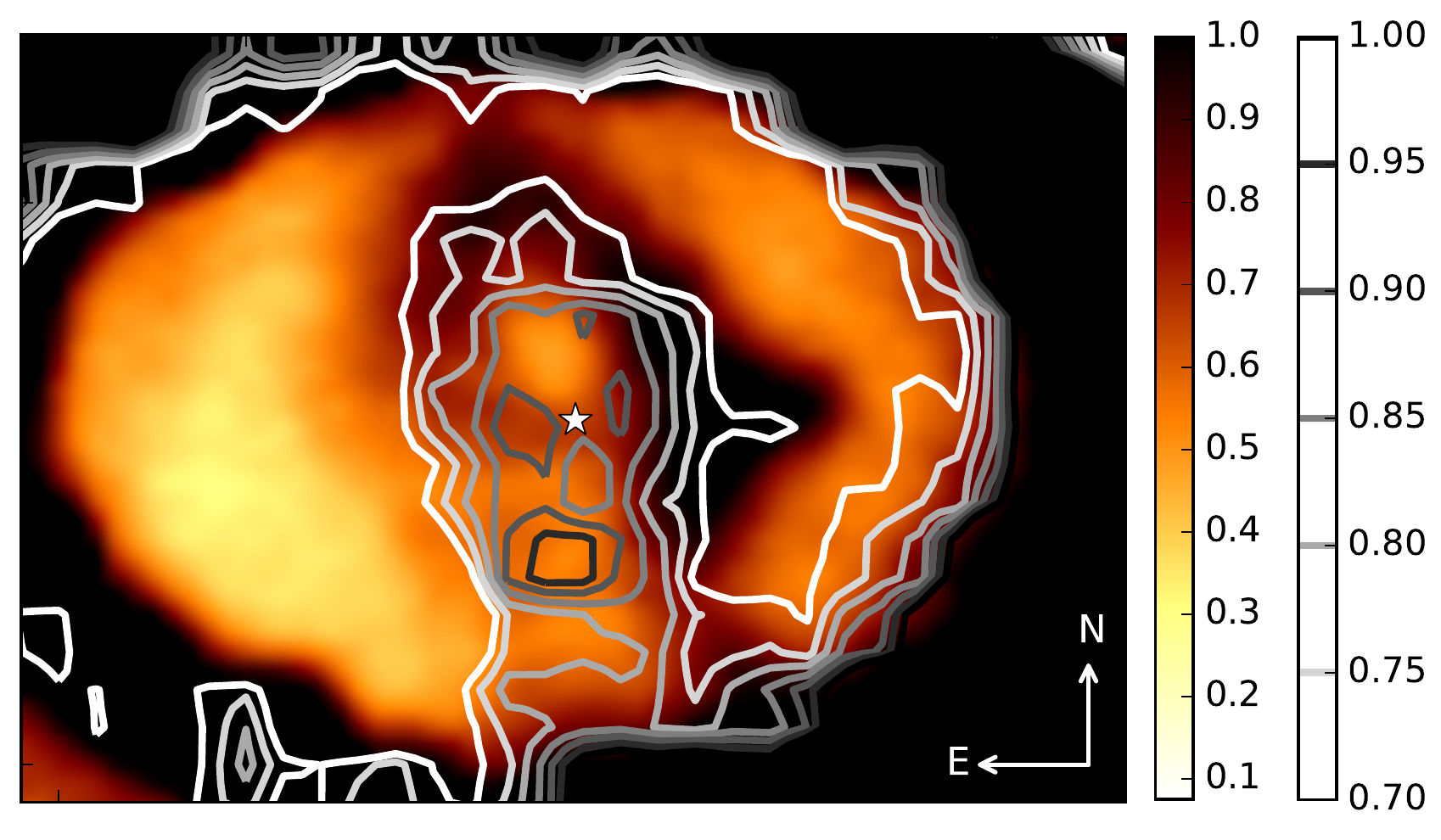}}
\caption{WFC3/F625W image from 2018 (color) together with the contours from the H$_2$~2.12~$\mu$m image (gray-scale, cf. left panel of Fig.~\ref{fig:h2conts}). The images have been normalized such that the maximum of the color scale has a value of 1. The F625W image is dominated by H$\alpha$ emission. The center of the ring is indicated by the white star. The FOV is $1\farcs{5} \times 1\farcs{0}$. 
 \label{fig:hst_h2}}
\end{figure}

In order to compare the distribution of H$_2$ with atomic H, we make use of HST observations that probe emission from H$\alpha$. Fig.~\ref{fig:hst_h2} shows the contours of the H$_2$ $2.12\ \mu$m emission (Fig.~\ref{fig:h2conts}) superposed on a recent WFC3/F625W image, taken about six months after the H$_2$ observations (Table \ref{tab:compobs}). The F625W filter is strongly dominated by H$\alpha$  (see Fig.~1 of \citealt{Larsson2013}). Fig.~\ref{fig:hst_h2} shows that the brightest H$_2$ emission is located further to the east than the brightest regions in H$\alpha$. In particular, the bright H$_2$ clump in the south coincides with a region of low surface brightness in the F625W image. The total extent of the H$_2$ emission is also smaller than that of H$\alpha$, as expected from the much higher velocities observed for the H$\alpha$ line (e.g., L16). 

We use the HST/STIS observation from 2014 (Table~\ref{tab:compobs}) in order to compare with the 3D distribution of H$\alpha$. In this observation the whole inner ejecta were covered by narrow slits (see L16 for details).  The H$\alpha$ 3D map has a spectral resolution of $450$~\kms, while the resolution in the plane of the sky is limited by the $0\farcs{1}$ slit width, which corresponds to 870 \kms\  at the time of the observation. We also note that the morphology of the ejecta in H$\alpha$ has not changed significantly in the 3.5 years between the H$\alpha$ and H$_2$ observations (cf. Fig.~\ref{fig:hst_h2} with Fig.~1 of L16). A direct comparison of the 3D velocity maps should therefore give a good view of the present-day H$\alpha$ and H$_2$ distributions. 

\begin{figure}[t]
\resizebox{90mm}{!}{\includegraphics{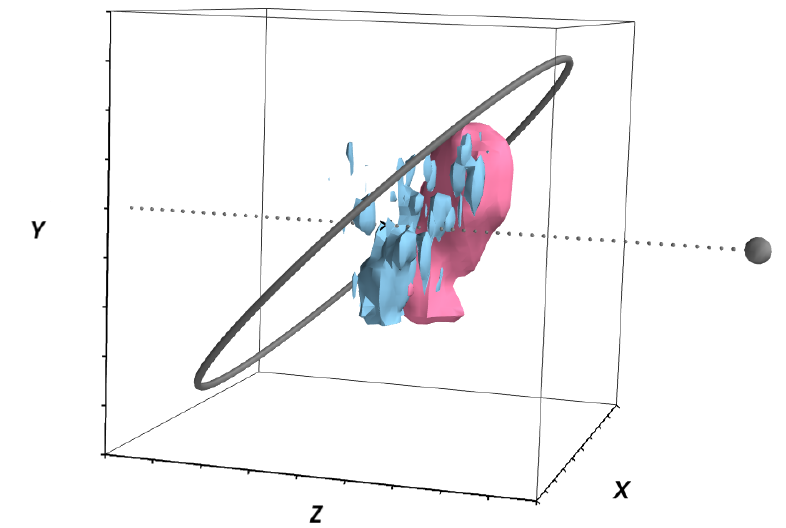}}
\caption{3D iso-surfaces of H$_2$ (blue) and H$\alpha$ (pink). The surfaces correspond to 45$\%$ of the maximal intensity for each line. The ring indicates the position of the reverse shock at the inner edge of the ring.  The dashed line and blob show the line-of-sight and position of the observer, respectively. The tick marks on the coordinate axis are separated by $1000$~\kms. An animated (rotating) version of this figure is available online. As the animation runs, three different semi-transparent contour levels are shown for H$\alpha$ (30, 45 and 65$\%$ of max). \label{fig:ha_h2_3d}}
\end{figure}

The result of the 3D comparison is shown in Fig~\ref{fig:ha_h2_3d}. An animated version is provided in the online material, showing different viewing angles as well as different contour levels for H$\alpha$. The comparison confirms the results from the images in Fig.~\ref{fig:hst_h2}, clearly showing that H$_2$ is located further to the east than H$\alpha$. In addition, the 3D maps reveal a similar distribution of H$\alpha$ and H$_2$ along the north-south direction, with a transition to gradually more redshifted emission when going from north to south (as quantified by the blue/red ratios in section \ref{sec:h23d}). The anti-correlation between the brightest regions of H$\alpha$ and H$_2$ clearly shows the effects of the X-ray emission from the ring, as discussed in Section \ref{sec:disc} below. 

\subsection{Comparison with ALMA observations of CO, SiO and dust}

ALMA has detected rotational transitions from a number of molecular species in the ejecta \citep{Matsuura2017}. The CO~2--1 and SiO~5--4 lines have also been mapped in 3D \citep{Abellan2017}. The ALMA data has a resolution of $\sim 100$~\kms\  along the line of sight and $0\farcs{05}$ in the plane of the sky, equivalent to 420~\kms\  at the time of the observation. The average epochs of the ALMA and H$_2$ observations are separated by only about two years. 

\begin{figure}[t]
\resizebox{90mm}{!}{\includegraphics{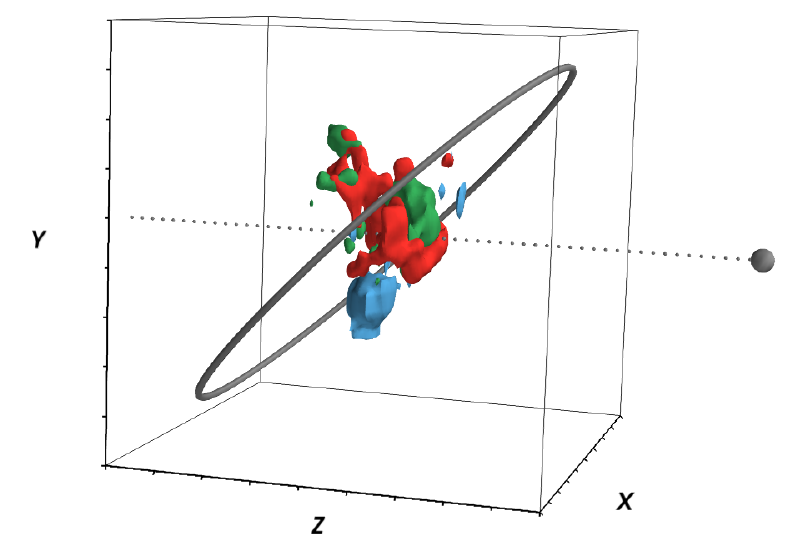}}
\caption{3D iso-surfaces of H$_2$ (blue), CO (red) and SiO (green). The surfaces correspond to 60$\%$ of the maximal intensity for each line. The ring indicates the position of the reverse shock at the inner edge of the ring.  The dashed line and blob show the line-of-sight and position of the observer, respectively. The tick marks on the coordinate axis are separated by $1000$~\kms. An animated (rotating) version of this figure is available online.  \label{fig:alma_h2_3d}}
\end{figure}

A comparison of iso-surfaces corresponding to  60$\%$ of max for CO, SiO and H$_2$ is shown in Fig.~\ref{fig:alma_h2_3d}. This shows that the brightest regions of molecular emission are clearly distinct. In particular, there is no strong CO and SiO emission overlapping with the bright clump of H$_2$ in the south. The brightest CO and SiO emission is instead distributed in ring/torus-like structures perpendicular to the plane of the ring. When considering fainter emission we note a small amount of overlap between the H$_2$ clump and the faint, southern tails of CO and SiO (shown in \citealt{Abellan2017}), as well as between weaker H$_2$ emission and the blueshifted structures of CO and SIO in the north. This overlap will at least partly be caused by the limited resolution. The relation between these fainter emission regions will be studied in detail together with HCO$^+$ in M. Matsuura et al. (in prep.). 

\begin{figure}[t]
\resizebox{90mm}{!}{\includegraphics{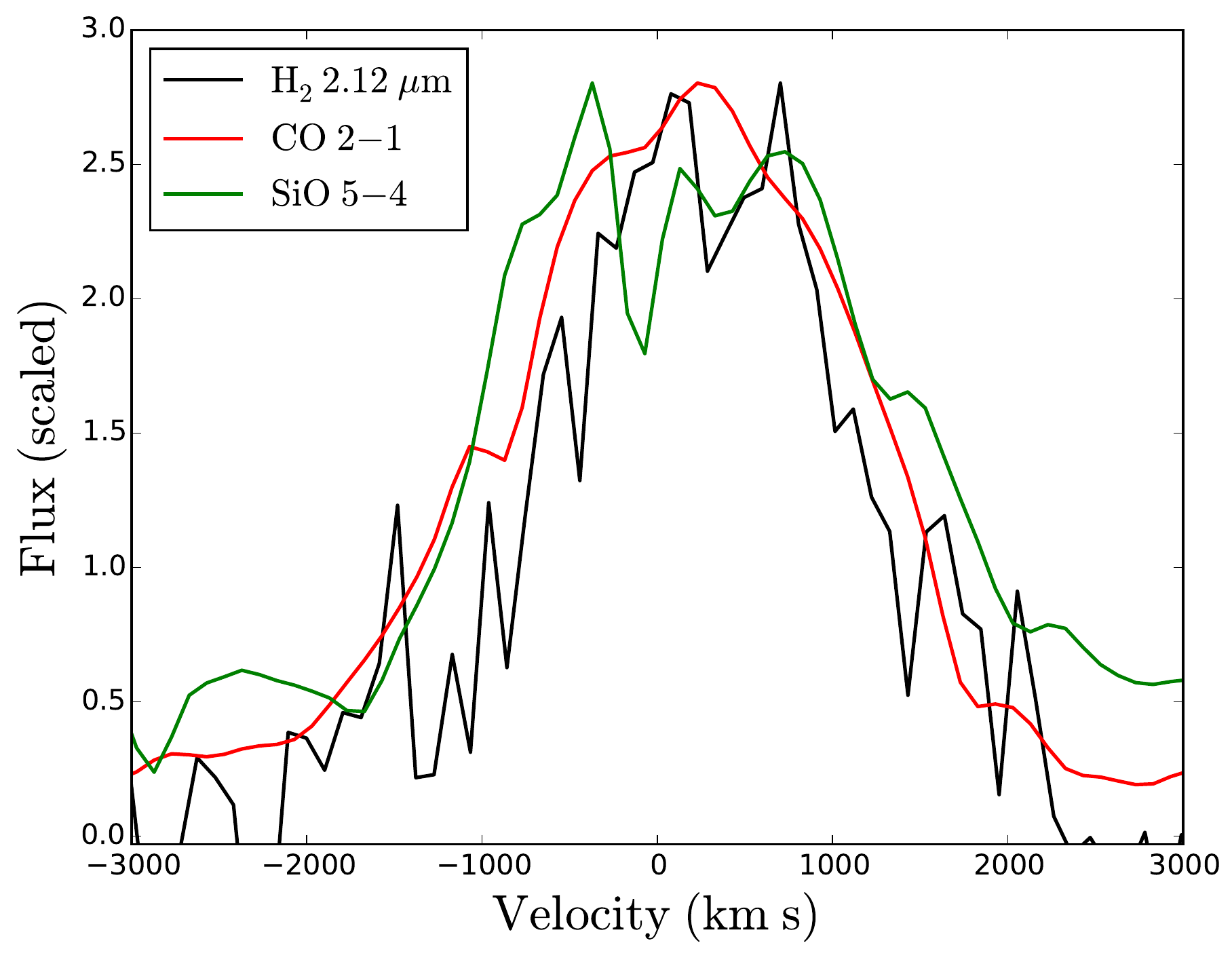}}
\caption{Integrated line profiles for the three molecular lines. The H$_2$ profile is the average of the SINFONI observations from 2005, 2007 and 2017. The spectrum has been binned by a factor of three for visual clarity. The CO and SiO profiles are from ALMA.  \label{fig:alma_h2_spec}}
\end{figure}

In order to compare the full velocity distributions of the three molecular lines, we also extract integrated line profiles from the ejecta. These are shown in Fig.~\ref{fig:alma_h2_spec}. In the case of H$_2$, the line is co-added from the observations with the best quality (2005, 2007 and 2017) to improve the S/N.  The individual spectra were extracted from an expanding aperture in order to account for the expansion of the ejecta between the observing dates. Our study of the time-evolution (see Section \ref{sec:time} below) shows that there is no significant spectral evolution between these observations. The profiles of the different molecular lines are clearly irregular, as expected from the 3D maps. This includes a dip a the centre of the SiO line, previously discussed in \cite{Matsuura2017}. As a crude estimate of the widths, we fit Gaussian lines, which give $\rm{FWHM_{CO}} = 2470 \pm 40$, $\rm{FWHM_{SiO}} = 3140 \pm 100$, and $\rm{FWHM_{H_2}} = 2100 \pm 80$~\kms. This shows that H$_2$ has a slightly smaller extent in velocity, as also seen directly in Fig.~\ref{fig:alma_h2_spec}.

\begin{figure}[t]
\resizebox{90mm}{!}{\includegraphics{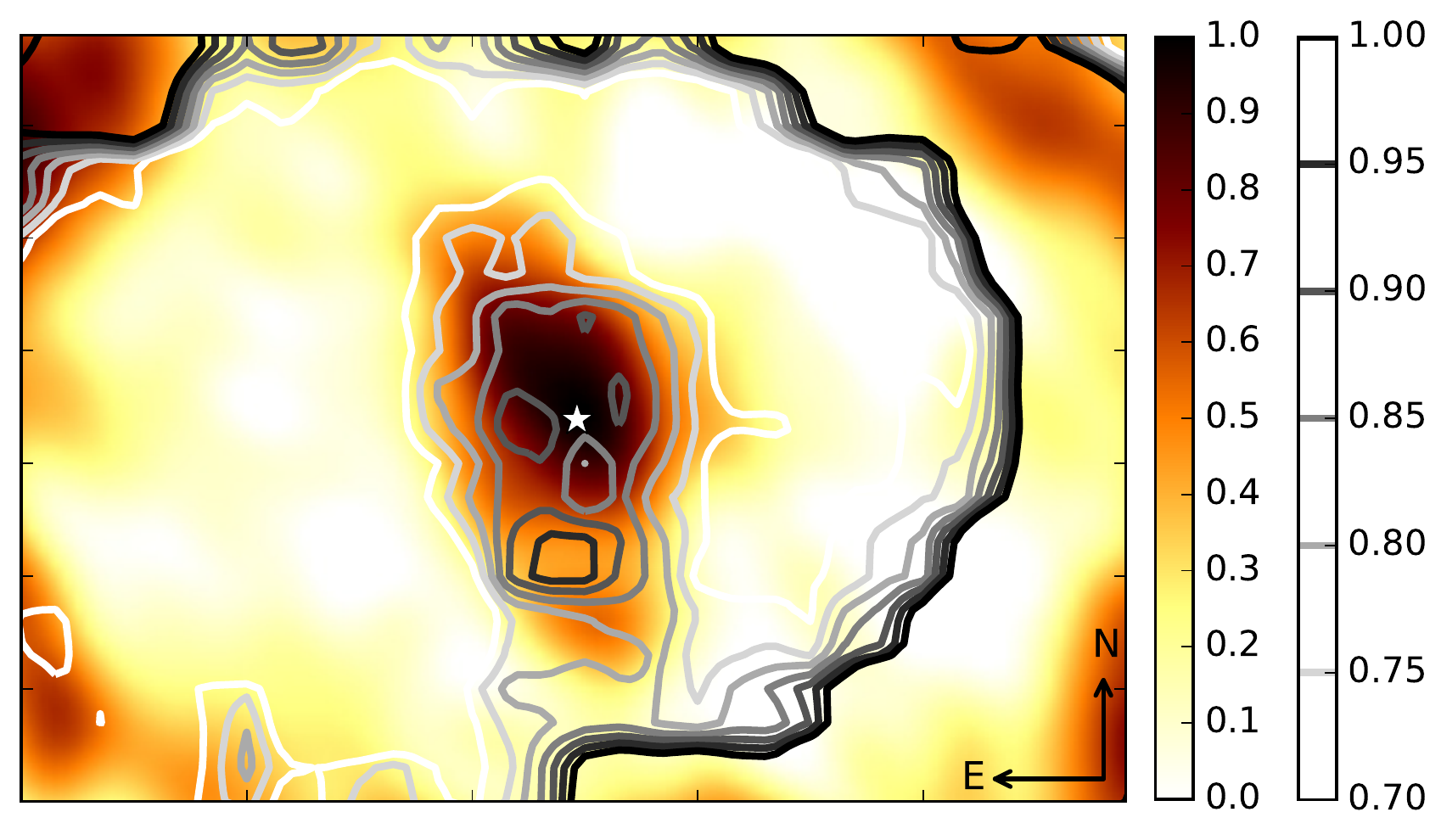}}
\caption{ALMA dust image at 315~GHz from 2015 (color) together with the contours from the H$_2$~2.12~$\mu$m image (gray-scale, cf. left panel of Fig.~\ref{fig:h2conts}). The images have been normalized such that the maximum of the color scale has a value of 1. The center of the ring is indicated by the white star. The FOV is $1\farcs{5} \times 1\farcs{0}$.  \label{fig:dust_h2}}
\end{figure}

The comparison of the different molecules may be affected by the dust in the ejecta \citep{Matsuura2015}. The dust is expected to obscure some of the H$_2$ emission from the far side of the ejecta, but not the CO and SiO. Fig.~\ref{fig:dust_h2}  shows the contours of the H$_2$ emission superposed on the ALMA dust image at 315~GHz. The ALMA observations were obtained approximately 2.5 years prior to the SINFONI observations of H$_2$, during which time the ejecta will have increased in size by only 8$\%$. As seen in Fig.~\ref{fig:dust_h2}, the dust emission is concentrated to the central ejecta, but there is also a fainter and smaller clump further to the south. It is notable that the peak of the H$_2$ emission is located in the gap in the dust image, between the main clump and the smaller clump. The dust clumps may be obscuring emission from any H$_2$ located behind or together with them. \\\\

\section{Time-evolution of H$_2$}
\label{sec:time}
\begin{figure*}[t]
\plotone{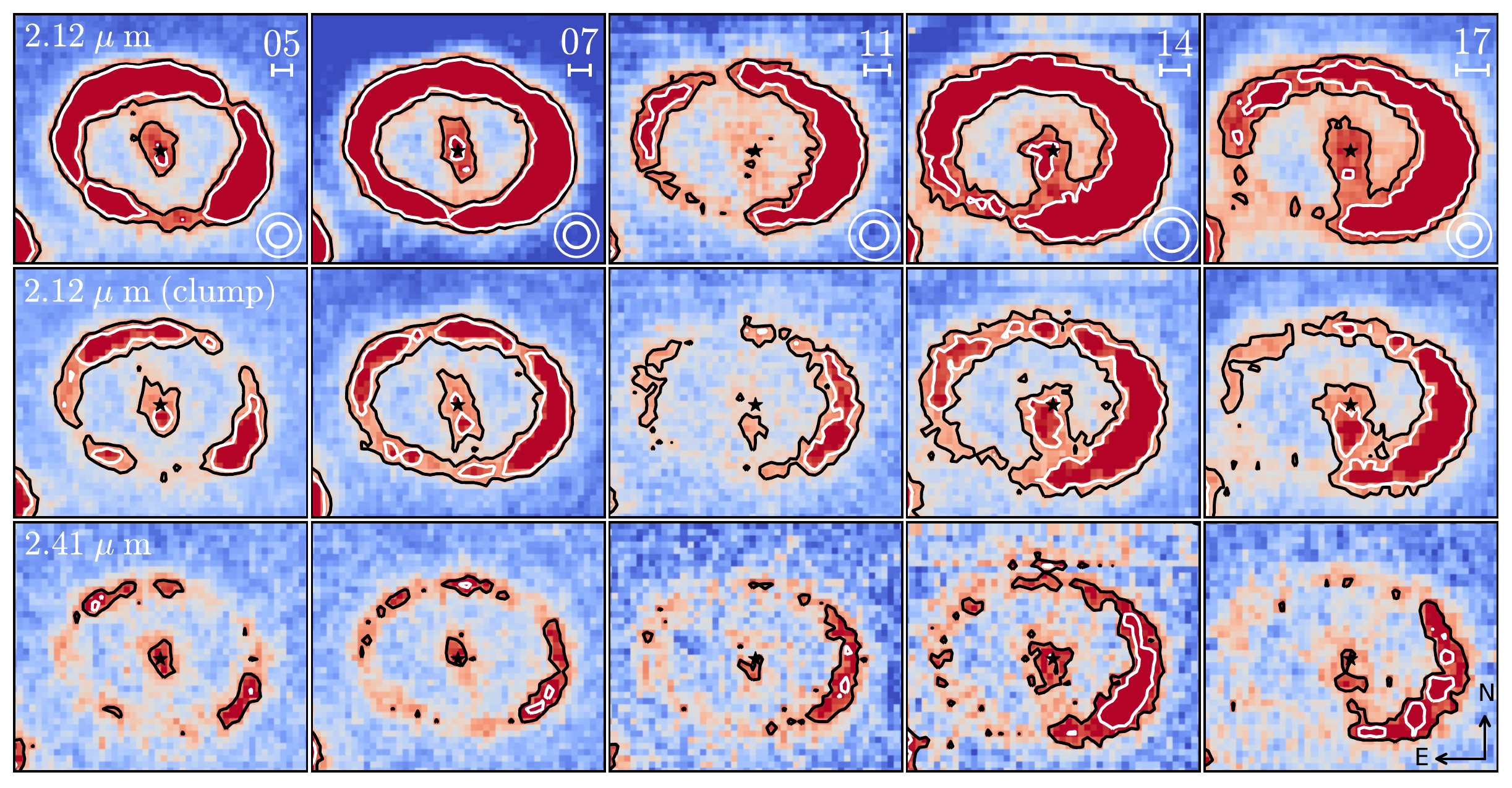}
\caption{Time evolution of the morphology of the H$_2$ emission. The top rows shows the $2.12\ \mu$m line inside $\pm 2000$~\kms, the middle row the $2.12\ \mu$m line in the interval $[-200,1000]$~\kms, and the bottom row the $2.41\ \mu$m line inside $\pm 2000$~\kms. The velocity interval in the middle row was selected to isolate the emission from the clump in the southern ejecta. The dynamic range for each of the images has been adjusted to show the ejecta morphology as clearly as possible. The black and white contours illustrate the 4 and 8$\sigma$ levels above the background inside the ring. The scale bars in the upper, right corners of the top row show the size of $1000$~\kms\  in the freely expanding ejecta for each epoch, while the circles in the lower, right corners show the size of $50\%$ and $80\%$ encircled energy of the PSF.  The FOV is $2\farcs{4} \times 2\farcs{0}$. \\ \label{fig:morphev}}
\end{figure*}

The time evolution of the flux and morphology of the H$_2$ emission provides information about the energy source ($^{44}$Ti or X-rays) as well as about possible formation/destruction of molecules with time. \cite{Fransson2016} measured the fluxes of the two H$_2$ lines between 2005--2014 and found no evidence for significant changes. Here, we only consider the flux in the $2.12\ \mu$m line due to the poor S/N around $2.41~\mu$m. Integrating over  $\pm 2000$~\kms, the flux in 2017 is $2.5\pm0.3 \times 10^{-16}$~\ergcms, which corresponds to a luminosity of $7.7\pm0.9\ \times 10^{31}\ \rm{erg\ s^{-1}}$. The error bar includes the statistical uncertainty as well as the $10\%$ uncertainty in the absolute flux calibration.

 The continuum level subtracted for the flux measurement was determined by fitting a straight line in the intervals $\pm (2000-3500)$~\kms. This is somewhat different from  the continuum interval used in \cite{Fransson2016}. When considering the previous fluxes reported in that paper, this change results in values that are  consistently $\sim 20\%$ higher, which gives an idea of the systematic uncertainty from the continuum subtraction. The continuum level varies with time, as discussed in Appendix \ref{app:cont}, and it makes up $\sim$~50--70$\%$ of the total flux inside $\pm 2000$~\kms\  in the different observations. When measuring fluxes from the previous observations we have accounted for the expansion of the ejecta by adjusting the size of the aperture (defined in Section \ref{sec:specim}). 

Given the uncertainties, the flux in 2017 is consistent with the fluxes measured in all the previous observations. However, we note some marginal ($\sim 2\sigma$) evidence for an increase with time, with the fluxes in 2014 and 2017 being higher than in 2005. The line profiles do not show any evidence for significant time evolution, with the line widths of all the observations being consistent with each other.  We thus conclude that there is no statistically significant time evolution of the H$_2$ emission in the spectra. 

The time-evolution of the morphology of the H$_2$ emission is shown in Fig.~\ref{fig:morphev}. In addition to images of the full $\pm 2000$~\kms\  intervals around the two lines, we also show images in the $[-200,1000]$~\kms\ interval for the $2.12\ \mu$m line, which isolates the emission from the bright clump in the south.  The comparison between the different epochs is complicated by large differences in data quality, both in terms of S/N and spatial resolution, as well as the expansion of the ejecta and strong evolution of the ring. To illustrate the varying spatial resolution we show the area corresponding to $50\%$ and $80\%$ encircled energy of the PSF for each epoch in Fig.~\ref{fig:morphev}. We also show a scale-bar corresponding to $1000$~\kms\ in the ejecta to demonstrate the effect of the expansion. In order to account for the varying S/N, we overlay contours on each image corresponding to the 4$\sigma$ and 8$\sigma$ levels above the background inside the ring, where $\sigma$ was measured in source-free regions on each image (cf.~Section \ref{sec:specim}). This shows that the 2011 observation has poor S/N and that the signal is always poor in the  $2.41\ \mu$m line. The lowest spatial resolution is in the observation from 2014. 

Given these uncertainties, we can draw two conclusions from Fig.~\ref{fig:morphev}. First, it is clear that the main clump of H$_2$ in the south is present in all observations. The middle row of Fig.~\ref{fig:morphev}  shows this clump moving to the south with time as the ejecta expand. Secondly, the full images of the $2.12\ \mu$m line show evidence for stronger emission in the north-east at early times and in the south at late times. This change can most likely be explained by the contribution from the continuum, as discussed in Appendix \ref{app:cont}. As in the case of the spectra, we therefore conclude that there is no statistically significant time evolution in the H$_2$ emission. However, we stress that we are limited by the data quality in several of the observations, and smaller changes in flux and/or morphology cannot be ruled out.

\section{Discussion}
\label{sec:disc}

\subsection{Formation, excitation and evolution of H$_2$ in SN~1987A }
 
Theoretical models for the formation of H$_2$ in SN~1987A were first presented by \cite{Culhane1995}. They found that H$_2$ forms mainly between 400 and 1000 days, after which the abundance freezes out at $\sim 1~\%$ of the total H abundance. With our SINFONI observations we probe the H$_2$ emission at much later times, between 2005 and 2017 ($\sim$ 6800--11,300 days). The time-evolution of H$_2$ at these epochs depends on the energy source that leads to excitation and/or destruction of the molecules. The two main energy sources currently operating in the ejecta are radioactive decay of $^{44}$Ti and  X-ray emission from the ring. The latter has had a strong impact on the optical emission from the ejecta since the year 2000 (5000 days), causing a significant brightening and a transition to an edge-brightened morphology \citep{Larsson2011,Larsson2013,Fransson2013}. Most recently, the energy input from X-rays is manifested in the brightening of a blob in the western part of the ejecta (L16, see also Fig.~\ref{fig:hst_h2}). 

During the approximate time period covered by the SINFONI observations, the soft (0.5--2~keV) X-ray flux has increased by a factor of $\sim 3$ (\citealt{Frank2016}, where the last reported X-ray flux is from the end of 2015). The X-rays can ionize and/or dissociate the H$_2$, resulting in a change of the H$_2$ flux. We note that even the destruction of H$_2$ can lead to an increase in flux since the molecules will deexcite if they re-form. Whether the increasing X-ray flux results in an increase or decrease of H$_2$ flux depends sensitively on the location of H$_2$ within the ejecta, as well as the intensity, spectrum and morphology of the X-ray emission from the ring.

The fact that no significant time evolution of H$_2$ was observed is instead consistent with the emission being powered by  the $^{44}$Ti decay, in agreement with the conclusions of \citealt{Fransson2016}. This scenario is also supported by the fact that the H$_2$ is located in the inner ejecta, since most of the X-ray emission is predicted to be absorbed further out \citep{Fransson2013}. The ejecta velocities where the X-rays are absorbed are not expected to change much with time. As an example, the model in \cite{Fransson2013} shows that the velocity corresponding to the peak of the energy deposition of 1~keV photons is 3390~\kms\ at 20 years and 3500~\kms\ at 30 years. 

The decay lines from $^{44}$Ti have been detected in SN~1987A \citep{Grebenev2012,Boggs2015}, but the emission is not spatially resolved. However, the [Si~{\scriptsize I}]+[Fe~{\scriptsize II}] line at 1.644~$\mu$m is expected to be a good proxy for the spatial distribution of $^{44}$Ti since Si, Fe and Ti are produced in similar burning zones. The only remnant where $^{44}$Ti has been spatially resolved is Cas A \citep{Grefenstette2014}. In this case the emission does not correlate with Fe or Si, but we note that this may be due to observational biases. In particular, the Fe and Si in the interior of the remnant (where most of the $^{44}$Ti is seen) may have densities and ionization states that do not produce observable emission. For SN~1987A, the time evolution of the 1.644~$\mu$m line strongly supports a scenario where it is powered by $^{44}$Ti (L16).  Images of the 1.644~$\mu$m line further show emission in the inner ejecta (\citealt{Kjaer2010}, L16), consistent with $^{44}$Ti powering the H$_2$. 

With $^{44}$Ti as the energy source, the H$_2$ can be excited by either non-thermal electrons or UV fluorescence. The spectral modeling in \cite{Fransson2016} favored the latter, though the results were not conclusive. In this scenario the UV photons are generated internally in the ejecta, for example as a result of He~{\scriptsize I} two-photon emission \citep{Culhane1995,Jerkstrand2011}. We note that thermal excitation of H$_2$ is ruled out by the low temperatures in the inner ejecta (below a few 100 K, \citealt{Jerkstrand2011,Kamenetzky2013,Matsuura2017}). By contrast, the excitation temperature of vibrational H$_2$ is nearly 6000\,K \citep{Field1966}. Future observations over a wider wavelength interval with the {\it James Webb Telescope} are needed to conclusively discriminate between the non-thermal and UV excitation scenarios.

The model by \cite{Culhane1995} showed that the H$_2$ formation  is dominated by gas-phase reactions involving H$^+$ and H$^-$, while the contribution from formation on dust grains is negligible. Consistent with this, the comparison with the ALMA dust image shows that the brightest H$_2$ emission coincides with a region of very weak dust emission. However, the comparison is complicated by the lack of 3D information for the dust and the fact that the dust may obscure H$_2$ emission (discussed further below).

\subsection{Ejecta geometry and mixing}

 We find that the H$_2$ emission in SN~1987A is confined to the core-region, inside $\lesssim 2500$~\kms. The 3D emissivity is dominated by a single clump in the southern ejecta, which has a central velocity of $\sim 1700\ $\kms\ and FWHM$\sim 1400$\ \kms. To the north of this clump there is weaker H$_2$ emission along the plane of the ring.  The large-scale geometry, including a gradual transition to increasingly redshifted emission when going from north to south, is similar to that observed in a number of other emission lines, including H$\alpha$ (L16). While this geometry will be affected by the large mass of dust residing in the ejecta \citep{Matsuura2015}, we note that observations suggest that the dust is likely located in optically thick clumps that affect the optical and NIR in the same way (see \citealt{Larsson2013}, L16 and references therein).  

In the comparison with dust it is interesting to note that the southern clump of H$_2$ is located between the two peaks in the dust emission. One possible explanation for this is that the brightest H$_2$ emission is due to less dust obscuration at this particular position. However, dust obscuration is expected to work similarly across the NIR, which means that similar effects should be seen in the [Fe~{\scriptsize II}]+[Si~{\scriptsize I}] $1.644\ \mu$m and He~{\scriptsize I}~$2.058\ \mu$m emission, contrary to  observations (L16). This shows that the H$_2$ morphology is unlikely to be strongly affected by dust, but instead reflect the location of molecule formation and excitation.

The 3D morphologies of the different emission components in the ejecta contain information about the progenitor structure and the explosion itself. The explosion geometry of SN~1987A shows some features that are qualitatively similar to neutrino-driven 3D explosion models, although there appears to be a higher degree of large-scale asymmetry in the observations (\citealt{Larsson2013}; L16). It should be noted that these comparisons are limited by the fact that the observed 3D emissivities do not only depend on the mass distributions, but also the relevant energy sources, dust obscuration and, in the case of H$_2$, the details of molecule formation/destruction. These effects need to be accounted for before any clear conclusions can be drawn regarding the agreement between models and observations. The ALMA observations of molecules are not affected by dust and a comparison with explosion models  was carried out in \cite{Abellan2017}, using the neutrino-driven 3D explosion models from \cite{Wongwathanarat2015}, continued until $\sim 150$~days. Some similarities with the models were found, but none of the progenitor models considered could reproduce all aspects of the observations. These comparisons may still be affected by the details of the ejecta chemistry, energy sources and disassociation of molecules. 

The comparison of H$_2$ with CO and SiO shows that the brightest regions of the different molecules occupy different parts of the inner ejecta. The FWHM of the integrated H$_2$ line profile is also slightly lower than for CO and SiO. The fact that we see H$_2$ in the innermost ejecta, at typical velocities slightly lower than for CO and SiO, provides direct evidence for mixing of the different nuclear burning zones of the progenitor star. Without such mixing, all the H$_2$ should be located further out than CO and SiO. Furthermore, the clumpy structures observed for all the molecules shows that Rayleigh Taylor instabilities at the time of the explosion and shortly thereafter broke up the gas and formed clumps.

The importance of mixing in SN~1987A was realized early on from the emergence of X- and $\gamma$-rays and the similarity of line profiles from different elements (see \citealt{McCray1993} for a review). It has also been seen in the previous 3D maps of SN~1987A (L16, \citealt{Abellan2017}) as well as in multi-dimensional explosion models \citep{Kifonidis2006,Hammer2010}. An important measure of the mixing is the lowest velocity of H. Previous observations have shown H$\alpha$ emission down to  $<700$~\kms\ (based on the integrated line profile,  \citealt{Kozma1998}) and $\sim 450~$\kms\ (based on narrow-slit spectroscopy, L16). 

From the SINFONI observations we detect significant H$_2$ emission even at zero velocity (i.e. at zero redshift at the centre of the SN). However, this is likely caused by the limited resolution, i.e. the scattered light in the tails of the PSFs. In order to investigate this, we created a simple model of the data cube, where we set all fluxes inside a radius of $400$~\kms\ in the image plane to zero. This corresponds to the central $2\times2$ pixels, where the centre as determined by \cite{Alp2018} is close to the intersection of these pixels. We further assume that there are point sources in all the immediately surrounding pixels (in the range $400-800$~\kms) with the same relative fluxes as in the real data. This model was convolved with the instrument resolution and normalized such that the pixels surrounding the empty region have the same total flux in the model and the data. We note that it is sufficient to consider the pixels immediately adjacent to the empty region, since adding point sources further out, while keeping the observed and model fluxes the same, would only serve to decrease the model flux at the centre. 

The PSF was modeled by a double Gaussian in order to account for the narrow core and broad wings. We determined the parameters of this model by fitting it to the star located to the south-west of the ring (usually referred to as star 3), which gives $\sigma_1 = 0\farcs{04} $, $\sigma_2 = 0\farcs{15}$  and a ratio of amplitudes $A_2/A_1 = 0.26$.  At the time of the observation the two values of $\sigma$ correspond to velocities of 315~\kms\ and 1180\ \kms, respectively. We also convolved the model with a Gaussian in the spectral direction, but note that this effect is completely negligible since the spectral resolution is so much better ($\sigma_s = 2.08$~\AA, equivalent to 29\ \kms). From this simple model we find that all emission detected inside $400$~\kms\  is consistent with being due to scattered light. However, creating the same kind of model, but with no emission inside 800~\kms, we find that the model under-predicts the observed flux by $\sim~30\%$. This indicates that H$_2$ is mixed into the range 400--800~\kms, consistent with the velocity constraints from H$\alpha$. 
 
The comparison of H$_2$ and CO is especially interesting in view of the detection of $\lesssim 5 \times 10^{-6}\ M_\sun$ of HCO$^+$ in the ejecta \citep{Matsuura2017}. The formation of HCO$^+$ requires H$_2$, which needs to either be co-located with CO or react with C and O atoms.  Our observations show that the brightest regions of H$_2$ and CO are clearly separated. However, when considering the weaker emission there is some overlap. We may also be missing some H$_2$ emission that is co-located with CO in the far side of the ejecta due to obscuration by dust. It is possible that Kelvin-Helmholtz instabilities between the H- and C/O-rich zones are sufficient for forming the small amount of HCO$^+$ observed. The presence of HCO$^+$ therefore does not contradict the conclusion of macroscopic mixing as the dominant mixing mode.  A detailed study of the distribution of HCO$^+$, including a comparison with CO and H$_2$, will be presented in a future work (M.~Matsuura et al., in preparation).

\section{Conclusions}
\label{sec:conclusions}

We have presented new SINFONI observations of the H$_2$ emission in the ejecta of SN 1987A, obtained 11,275 days after the explosion at the end of 2017. These observations provide information about the 3D emissivity of H$_2$.  We have compared the observations with previous observations of H$_2$, as well as with 3D maps of other emission lines in the ejecta. The main conclusions are summarized below. 

\begin{itemize}

\item The 3D geometry of H$_2$ is dominated by a single clump in the southern ejecta. The clump has a space velocity of $\sim 1700\ $\kms\ and FWHM~$\sim 1400$\ \kms. To the north of this clump there is weaker H$_2$ emission along the plane of the ring. We find that  the lowest velocities of H$_2$ are in range 400--800\ \kms, while the highest velocities are around 2500~\kms.   

\item No significant time-evolution of H$_2$ is observed between 6800 and 11,3000 days. The constant flux, together with the observed location of H$_2$ within the ejecta, is consistent with $^{44}$Ti being the dominant energy source powering the emission.

\item The H$_2$ follows a similar large-scale geometry as H$\alpha$, but is located further to the east. In particular, the brightest regions of H$_2$ coincide with weak H$\alpha$ emission, which is explained by the latter being powered by X-ray emission from the ring. 

\item A comparison of H$_2$ with CO and SiO shows that the brightest regions of the different molecules occupy different parts of the inner ejecta and that the FWHM of H$_2$ is slightly lower than for CO and SiO. These observations reflect the different excitation mechanisms for the molecules as well as the mixing in the explosion. 

\item The brightest H$_2$ emission coincides with a region of very weak emission in the ALMA dust image at 315~GHz. This is consistent with theoretical predictions that the H$_2$ in the ejecta should form in the gas phase rather than on dust grains. 

\end{itemize}

\acknowledgments

This work was supported by the Knut and Alice Wallenberg Foundation and the Swedish Research Council.  MM is supported by an STFC Ernest Rutherford fellowship (ST/L003597/1). HLG and PC acknowledge support from the European Research Council (ERC) in the form of Consolidator Grant CosmicDust (ERC-2014-CoG-647939). The SINFONI observations were collected at the European Organization for Astronomical Research in the Southern Hemisphere, Chile (ESO Programs 076.D-0558(A), 080.D-0727(C), 086.D-0713(C), 094.D-0505(C) and 0100.D-0705(C)). Support for HST GO programs 13401 and 15256 was provided by NASA through grants from the Space Telescope Science Institute, which is operated by the Association of Universities for Research in Astronomy, Inc., under NASA contract NAS5-26555. This paper makes use of the following ALMA data: ADS/JAO.ALMA\#2013.1.00063.S, 2013.1.00280.S and 2015.1.00631.S.
ALMA is a partnership of ESO (representing its member states), NSF (USA) and NINS (Japan), together with NRC (Canada), NSC and ASIAA (Taiwan), and KASI (Republic of Korea), in cooperation with the Republic of Chile. The Joint ALMA Observatory is operated by ESO, AUI/NRAO and NAOJ.

%

\facilities{VLT (SINFONI), HST(WFC3), ALMA}


\software{CASA (v4.5.1, \citealt{McMullin2007}),
	DrizzlePac \citep{Gonzaga2012},
	matplotlib \citep{Hunter2007},
	Mayavi \citep{Ramachandran2011}}





\appendix

\section{Investigation of the K-band continuum}
\label{app:cont}

\begin{figure*}[t]
\plotone{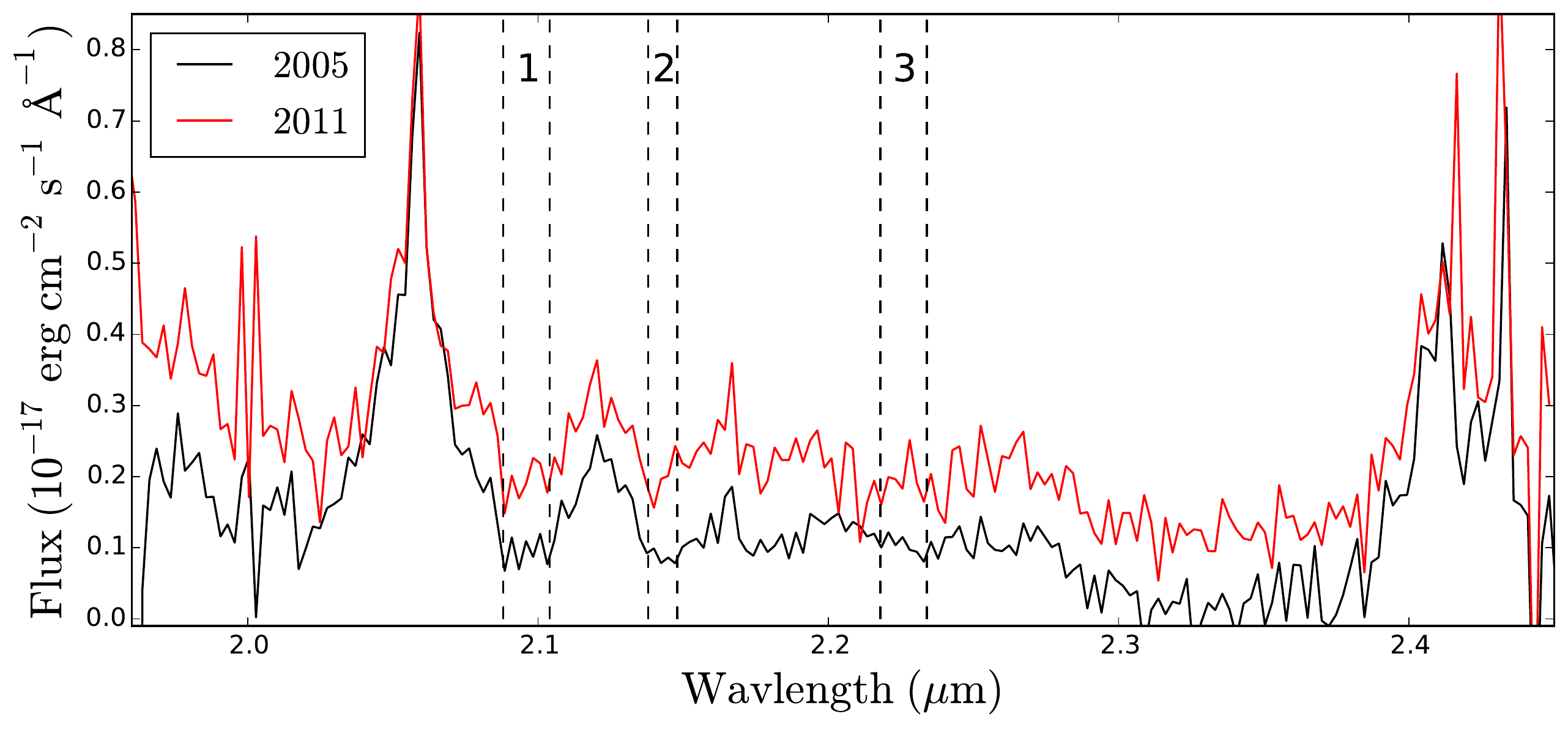}
\caption{SINFONI K-band spectra from 2005 (black) and 2011 (red), illustrating the increase in the continuum level. Both spectra have been binned by a factor of ten for visual clarity. The dashed lines indicate the three continuum intervals that were used for the flux measurements in Fig.~\ref{fig:contflux} and images in Fig.~\ref{fig:contims}. \\ \label{fig:contspec}}
\end{figure*}

\begin{figure}[t]
\begin{center}
\resizebox{80mm}{!}{\includegraphics{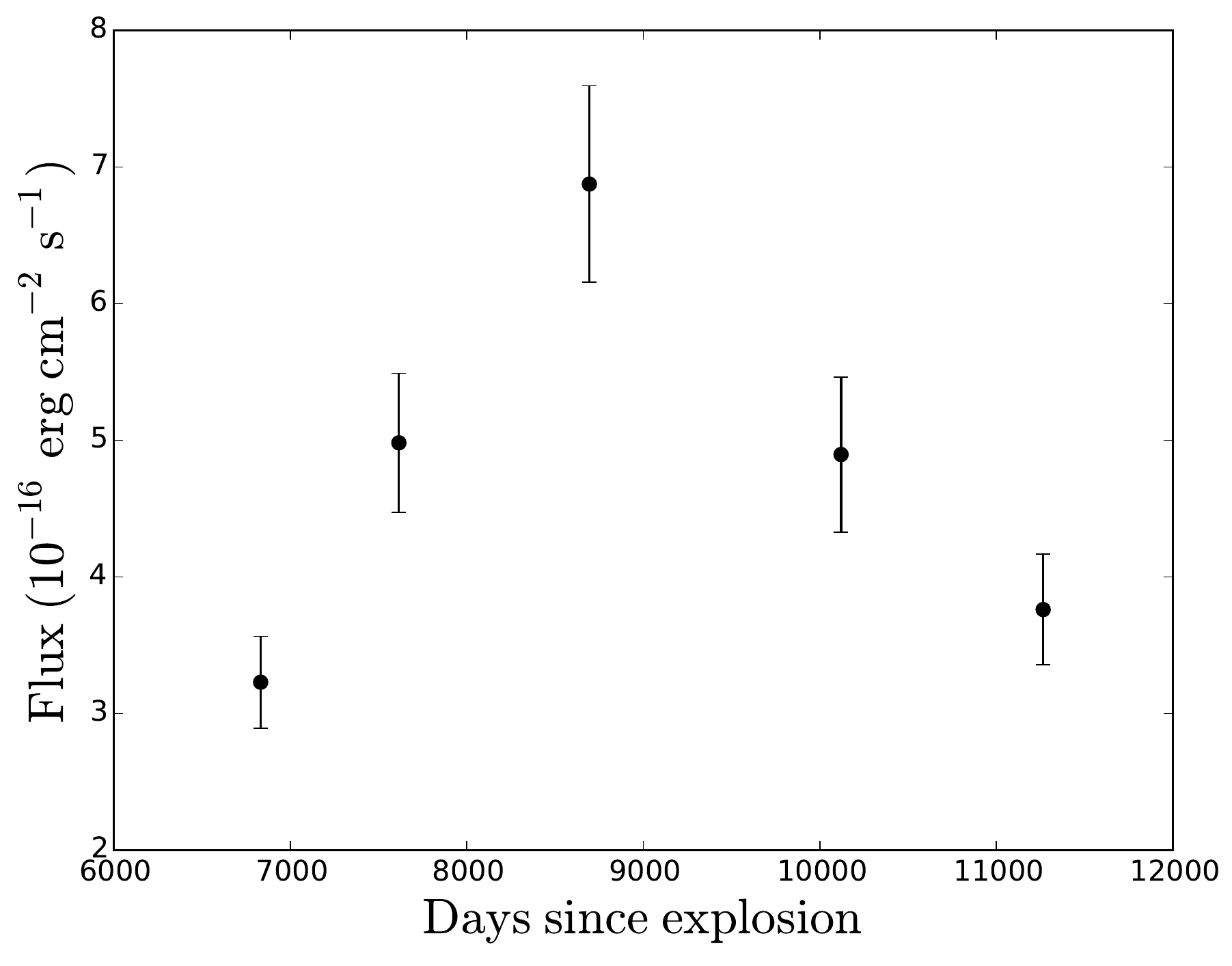}}
\end{center}
\caption{Temporal evolution of the sum of the flux in the three continuum intervals defined in Fig.~\ref{fig:contspec}.   \label{fig:contflux}}
\end{figure}

The K-band spectrum of the ejecta contains a significant continuum component in addition to the bright emission lines. At least part of this continuum is expected to be due to weak lines, mainly from He I and Fe II, as well as Al~I 2.1093 2.1163 $\mu$m near the H$_2$ line \citep{Kjaer2010,Fransson2016}. However, as seen in \cite{Fransson2016}, this is not sufficient to fully explain the observed flux. Here, we investigate the properties of the continuum and discuss its origin and the implications for our study of the H$_2$ emission. A study of the time-evolution of the spectra (see also Section \ref{sec:time}) reveals changes in the continuum level. This is illustrated in Fig.~\ref{fig:contspec}, where we show the K-band spectra of the ejecta  for the two observations with the lowest  and highest  continuum levels; from 2005 and 2011, respectively. The spectra from the other epochs are omitted for visual clarity. These can instead be found in \cite{Fransson2016}  and in Fig.~\ref{fig:kspec}. All spectra were extracted from the same expanding elliptical aperture as in Section~\ref{sec:time}. To quantify the changes in flux and morphology of the continuum, we identify three spectral regions (indicated by the dashed lines in Fig.~\ref{fig:contspec}) that are free of strong emission lines.  Fig.~\ref{fig:contflux} shows the total flux in these three regions as a function of time. There is an increase by a factor of two from 2005 to 2011, followed by a similar decrease until the last observation in 2017. However, the flux changes have relatively low statistical significance, with only the change between 2005 and 2011 being significant at the $3\sigma$ level. We find no evidence for changes in the shape of the continuum. 

\begin{figure*}[t]
\begin{center}
\resizebox{120mm}{!}{\includegraphics{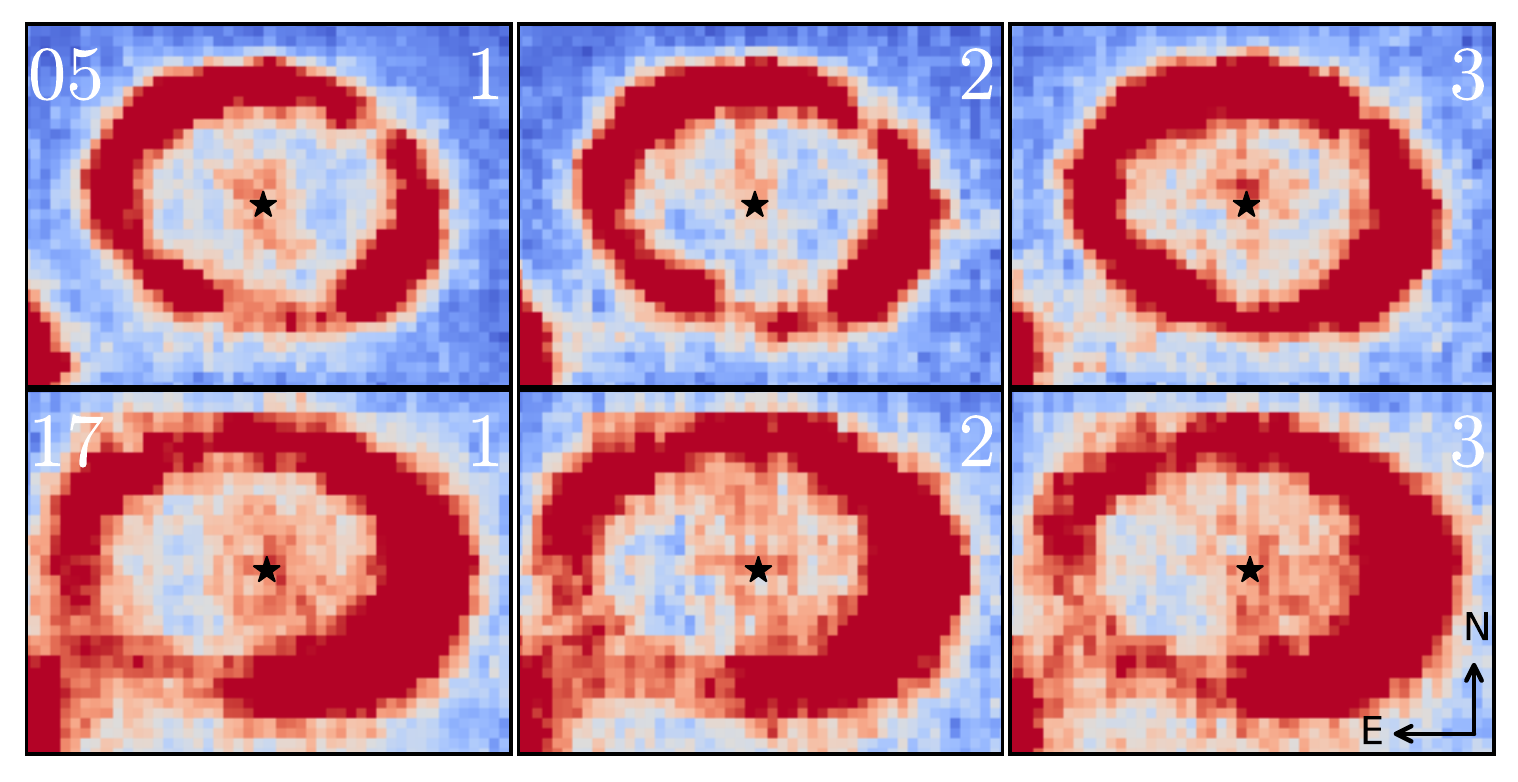}}
\end{center}
\caption{Images extracted from the three continuum intervals defined in Fig.~\ref{fig:contspec}. Observations from 2005 and 2017 are shown in the top and bottom rows, respectively. The dynamic ranges of the images were optimized to highlight the morphology of the ejecta. It is different for the three intervals, but the same for the two epochs in each of the intervals. Note the change in morphology, with the brightest region of the ejecta moving from the north-east in 2005 to the south-west in 2017. The FOV is $2\farcs{4} \times 1\farcs{8}$.  \label{fig:contims}}
\end{figure*}

Fig.~\ref{fig:contims} shows images from 2005 and 2017 extracted from the three continuum intervals. These two epochs were chosen because they have the best spatial resolution. The images reveal a clear systematic trend with time, with the brightest region of the ejecta being located in the north-east in 2005 and in the south-west in 2017. This change suggests that the continuum may be powered by the X-ray emission from the ring. {\it Chandra} images show that the strongest X-ray emission in the ring was in the north-east in 2005, while the west side started dominating around 2009 and was more than $50\%$ brighter than the east in 2016 \citep{Frank2016}. The increase in the continuum flux between 2005 and 2011 (Fig.~\ref{fig:contflux}) is also consistent with the increasing X-ray flux during the same time period \citep{Frank2016}. However, the X-ray light curve has leveled off since 2012, while we see a decrease in the continuum flux between 2011 and 2017 (significant at $2.5\sigma$). This may indicate that extreme UV/soft X-ray emission below the {\it Chandra} energy range dominates the energy input to the ejecta (as also discussed by \citealt{Fransson2013}). This emission is expected to correlate more closely with the optical light curve of the ring, which has been decreasing since 2009 \citep{Fransson2015}. 

Two possibilities for a true continuum component in the K-band were discussed in \cite{Fransson2016}; high-latitude synchrotron emission from the ring and emission from ultra-small dust grains in the ejecta. The former should give bright emission near the ring, particularly in the east where this component is strong (e.g., \citealt{Indebetouw2014}), while the latter should originate from the centre of the ejecta and not change significantly with time. The images in Fig.~\ref{fig:contims} do not agree with either option. Instead, Fig.~\ref{fig:contims} shows that the emission is associated with the ejecta, but at high enough velocities to be affected by the X-rays from the ring. We therefore propose that the continuum is due to a mixture of the weak atomic lines predicted by the model and the high-velocity wings from the strongest lines. The line profiles in the model by \cite{Fransson2016} may be somewhat too narrow, as suggested by inspection of their Fig.~3. The velocity of the He I 2.058~$\mu$m line at the blue wing of the H$_2$ 2.12~$\mu$m line is $\sim 7000$~\kms, while the velocity of Br$\gamma$ is $\sim -4000$~\kms\ at the red wing. Such high velocities are clearly seen in other lines (in particular in H$\alpha$), making such a contribution plausible. There may also be small contributions from the reverse shock, where velocities extend up to $> 10,000$~\kms\ (e.g., \citealt{France2011}), other weak H$_2$ lines (in the scenario of UV excitation, \citealt{Fransson2016}) and scattering of ring emission by the dust. 

The continuum emission will have some, relatively minor, effects on our study of the morphology of the H$_2$ emission at $2.12\ \mu$m. The most significant effect is that images of the line will show some excess emission to the north-east at early times and to the south-west at late times (as in Fig.~\ref{fig:contims}). However, the brightest regions of the H$_2$ images are almost a factor of three above the continuum and will not be significantly affected.  We note that the continuum cannot simply be subtracted from the images since its morphology changes somewhat with wavelength, as also seen in Fig.~\ref{fig:contims} and expected if it is due to line emission. Finally, our 3D iso-surfaces of H$_2$ in Section \ref{sec:h23d} will not be affected  since  we only consider emission levels well above the continuum.




\end{document}